# Monomer-resolved simulations of cluster-forming dendrimers

Dominic A. Lenz,[†] Bianca M. Mladek,[‡] Christos N. Likos,[¶] Gerhard Kahl,[§] and Ronald Blaak[*,†]

Institute of Theoretical Physics, Heinrich-Heine-Universität Düsseldorf, Universitätsstraße 1, D-40225 Düsseldorf, Germany, Department of Chemistry, University of Cambridge, Lensfield Road, Cambridge CB2 1EW, United Kingdom, Faculty of Physics, University of Vienna, Boltzmanngasse 5, A-1090 Vienna, Austria, and Center for Computational Material Science and Institute of Theoretical Physics, Vienna University of Technology, Wiedner Hauptstraße 8-10, A-1040 Vienna, Austria

E-mail: blaak@thphy.uni-duesseldorf.de

---

[*]To whom correspondence should be addressed
[†]Institute of Theoretical Physics, Heinrich-Heine-Universität Düsseldorf, Universitätsstraße 1, D-40225 Düsseldorf, Germany
[‡]Department of Chemistry, University of Cambridge, Lensfield Road, Cambridge CB2 1EW, United Kingdom
[¶]Faculty of Physics, University of Vienna, Boltzmanngasse 5, A-1090 Vienna, Austria
[§]Center for Computational Material Science and Institute of Theoretical Physics, Vienna University of Technology, Wiedner Hauptstraße 8-10, A-1040 Vienna, Austria

September 24, 2010

Abstract

We present results of monomer-resolved Monte Carlo simulations for a system of amphiphilic dendrimers of second generation. Our investigations validate a coarse-grained level description based on the zero-density limit effective pair-interactions for low and intermediate densities, which predicted the formation of stable, finite aggregates in the fluid phase. Indeed, we find that these systems form a homogeneous fluid for low densities, which on increasing the density spontaneously transforms into a fluid of clusters of dendrimers. Although these clusters are roughly spherical in nature for intermediate densities, also more complex structures are detected for the highest densities considered.

# 1 Introduction

The analysis of cluster formation in complex fluids is a topic that has attracted considerable attention in recent theoretical studies.[1–6] The possibilities and conditions of generating cluster phases—i.e. an equilibrium state in which the centers of mass of particles or molecules are able to lie on top of each other and thus build stable structures—have been extensively discussed over the last years. It has been shown that the cluster forming ability of a complex fluid depends decisively on the shape of the (effective) pair-interaction between particles or molecules. Bounded interactions provide a basis for full particle overlaps, and can therefore allow for particle clustering; counter-intuitively, this can even be the case for fully repulsive interactions as long as the sufficient condition is fulfilled that the potential is a member of the so-called $Q^{\pm}$-class of interactions, i.e., a class of potentials with negative components in their Fourier transform.[4,7,8]

Such kinds of interactions have been established for a few types of macromolecules in a coarse grained level description.[9,10] In such a treatment, the large number of internal degrees of freedom is averaged out, resulting in a spherically-symmetric effective pair-interaction. Within these studies, it was found that dendrimers, which represent a class of synthetic macromolecules characterized by well described tree-like architectures,[11,12] are a promising candidate for cluster formation.[9] Due to their enormous potential and recent advances in chemical techniques that allow for tailoring well characterized molecules, dendrimers already have been in the spotlight of various branches of science in both experimental and theoretical studies .[13–15]

The basic architecture of a dendrimer consists of monomeric units that have a minimum functionality of three that enables each unit to chemically bind to three or more other units. By starting with a single or a pair of these units, it is possible to grow regularly branched macromolecules in a controlled fashion, generation after generation, by chemically attaching new units to the available binding sites. This finally brings about the tree-like hybrid structure of dendrimers that lies somewhere in between polymer chains and colloidal particles. The broad range of possible spatial conformations of dendrimers encompasses dense-core,[12,16–18] dense-shell,[9,16] or core-shell structures; this flexibility results in a large area of possible applications, which include gene transduction,[19] drug delivery,[20–22] bio-sensors,[23] contrast agents[13,24] and many more.

From a more fundamental point of view, recent theoretical studies have shown that the possibility to approximate dendrimers by soft spheres[9,25–27] leads to an excellent model system that can be used for the study of the cluster formation introduced above.[1,4,7,28] For the particular case of amphiphilic dendrimers, in which the macromolecules consist of a solvophobic core and a solvophilic shell, the effective pair-interaction belongs to the above mentioned $Q^{\pm}$-class of bounded potentials. Therefore, it is expected that such macromolecules possess the counter-intuitive ability to self-assemble in clusters at sufficiently high densities. While this behavior was corroborated by computer simulations applying a coarse grained level description,[1,5,9] it still requires a more careful analysis on the more fundamental level of its monomeric units. Indeed, the guidance offered by zero-density, pairwise-additive interactions is not always reliable: also ring polymers have been shown to interact by means of $Q^{\pm}$-potentials; however, at the finite densities at which clusters are predicted to form on the basis of pair interactions alone, many-body potentials come into play. The latter act in a way that is detrimental to aggregation of ring-shaped macromolecules, an effect that can be attributed to the shrinkage of the rings above their overlap concentration.[10]

In the present work, we show with the aid of Monte Carlo (MC) simulations[29] that the coarse grained approach using effective interactions between amphiphilic dendrimers in the zero-density limit forms an appropriate, semi-quantitative description of the system's behavior on a monomer-resolved level. In addition, the predicted cluster-forming ability is confirmed for these systems which spontaneously form stable, finite aggregates at sufficiently high densities.

The remainder of this paper is organized as follows: in Section 2 we highlight the main features of the dendrimer model that is used and summarize some of the former results and motivations. In Section 3 the cluster assignment and characterization method is outlined. The structural analysis of these systems is presented in Section 4. In Section 5 the internal structure of clusters is examined, and we finish in Section 6 with a short summary and concluding remarks.

## 2 Model and Systems

In order to investigate the cluster-formation of dendrimers in MC simulations, we adopt the model for second-generation, amphiphilic dendrimers put forward by Mladek et al.[9,30,31] Here, dendrimers are considered that consist of two central units, to which monomers with functionality f = 3 are attached. In total we have N = 14 monomers which—according to their interactions with the solvent—are divided into two classes, generating thereby the amphiphilic behavior. The two central monomers and the four monomers of the subsequent first generation form the solvophobic core particles, labeled by subscript C. The remaining eight exterior monomers form the solvophilic shell of the dendrimer, labeled by subscript S.

The bonds between adjacent monomers with a relative distance r are modeled by the finitely extensible nonlinear elastic (FENE) potential,

$$\beta \Phi_{\mu\nu}^{\text{FENE}}(r) = -K_{\mu\nu} R_{\mu\nu}^2 \ln\left[1 - \left(\frac{r - l_{\mu\nu}}{R_{\mu\nu}}\right)^2\right], \quad \mu\nu = \text{CC}, \text{CS}, \qquad (1)$$

where $\beta = 1/k_B T$ is the inverse temperature, the $K_{\mu\nu}$ are spring constants, and the $R_{\mu\nu}$ stand for the maximum deviation from the equilibrium bond lengths $l_{\mu\nu}$. All other interactions between two monomers separated by a distance r are modeled by the Morse potential,

$$\beta \Phi^{\text{Morse}}_{\mu\nu}(r) = \varepsilon_{\mu\nu} \left\{ \left[ e^{-\alpha_{\mu\nu}(r-\sigma_{\mu\nu})} - 1 \right]^2 - 1 \right\}, \quad \mu\nu = \text{CC}, \text{CS}, \text{SS}, \qquad (2)$$

which is characterized by a repulsive core at short and an attractive tail at longer distances. The depth and range of these potentials are parameterized by $\varepsilon_{\mu\nu}$ and $\alpha_{\mu\nu}$ respectively, and $\sigma_{\mu\nu}$ are the monomer diameters. The set of parameters used for the simulation corresponds to the $D_2$-model introduced by Mladek et al.[9] For completeness, they are summarized in Table 2, expressed in terms of the diameter of the core monomers $\sigma \equiv \sigma_{\text{CC}}$, which from now on designates the reference length scale. An effective measure for the size of a dendrimer is its radius of gyration $R_g$, which in the case of a single, isolated dendrimer of the type used here is given by $R_g \approx 3.36\sigma$. In contrast to the core-monomer diameter $\sigma$, the radius of gyration however has not a fixed value but depends on the overall density and structure of the system.

Table 1: Overview of the monomer interaction potential parameters used between core (C) and/or shell (S) monomers.

| Morse | $\varepsilon_{\mu\nu}$ | $\alpha_{\mu\nu}\sigma$ | $\sigma_{\mu\nu}/\sigma$ |
|---|---|---|---|
| CC | 0.714 | 6.4 | 1 |
| CS | 0.014 | 19.2 | 1.25 |
| SS | 0.014 | 19.2 | 1.5 |
| FENE | $K_{\mu\nu}\sigma^2$ | $l_{\mu\nu}/\sigma$ | $R_{\mu\nu}/\sigma$ |
| CC | 40 | 1.875 | 0.375 |
| CS | 30 | 3.75 | 0.75 |

Figure 1a shows the effective dendrimer-dendrimer interaction potential as a function of the distance between the centers of mass for the chosen $D_2$ dendrimer. This interaction is obtained by simulating two free dendrimers with the help of umbrella sampling,[9,30] representing therefore the effective pair-interaction in the infinite dilution limit. This potential can be fitted to a superposition of two Gaussian interactions of opposite sign,[9] namely

$$\beta \Phi_{\text{eff}}(r) = \varepsilon_1 \exp[-(r/R_1)^2] - \varepsilon_2 \exp[-(r/R_2)^2], \qquad (3)$$

with $\varepsilon_1 = 23.6$, $\varepsilon_2 = 22.5$, $R_1 = 3.75\sigma$, and $R_2 = 3.56\sigma$. The Fourier transform of the effective interaction $\beta\hat{\Phi}_{\text{eff}}(q)$ is shown in Figure 1b and reveals a negative range that arises from both the steepness of the interaction and the local minimum in the effective interaction near r = 0. According to the criteria put forward in Ref. Likos et al.[7], a system interacting via an interaction with such negative Fourier components is able to form clusters. In Ref. Likos et al.[4], it further has been analytically established that the inherent instability arising from the negative Fourier components of the pair potential $\Phi(r)$ drives the formation of clusters in the fluid phase at finite densities and the subsequent crystallization of the liquid into a cluster solid. It is not possible to give a precise value of the fluid density at which clusters in the liquid phase form (this would be the analog of the critical micelle concentration encountered, e.g., in block copolymer solutions), since cluster formation is, in our case, a gradual process. However, there exists an accurate prediction for the density $\rho_\times$ at which the cluster fluid undergoes a phase transition into a cluster crystal, namely[4]

$$\rho_\times \sigma^3 = \left[1.393\beta|\hat{\Phi}_{\text{eff}}(q_*)|/\sigma^3\right]^{-1}, \qquad (4)$$

where $q_*$ is the value of the wavenumber q at which $\hat{\Phi}_{\text{eff}}(q)$ attains its most negative value. Using the value for $\beta\hat{\Phi}(q_*)/\sigma^3 \cong 35$ (see Figure 1b), we obtain the estimate $\rho_\times \sigma^3 \cong 0.02$. Even though this density corresponds to a system that is still very dilute on the monomer level, it already exceeds the dendrimer overlap density $\rho_* \cong 3/(4\pi R_g^3) \cong 0.006$. Accordingly, one should expect that the interaction between dendrimers will significantly deviate from its zero-density limit, and hence the result for $\rho_\times$ thus obtained can only be used as a rough estimate of the range of densities for which clusters should exist and to orient us in our search for such phenomena in the full, monomer-resolved simulations. Therefore, it is a priori not known at what density our dendrimers will start to form clusters.

In our investigations we have prepared multiple, independent random configurations of N = 250 dendrimers at various densities in a cubic simulation box with the usual periodic boundary conditions. These configurations have been allowed to equilibrate for sufficiently long times to guarantee a structural relaxation by employing Monte Carlo simulations at constant density with local monomer displacements as well as full dendrimer translations and rotations. In the remainder of this paper we will focus on the analysis of only four systems, labeled A, B, C, and D, spanning a range of concentrations around the above-mentioned value $\rho_\times \sigma^3 = 0.02$ for which cluster crystals should appear.[4] In addition, systems at various intermediate densities have been simulated, but will not be discussed in detail here.

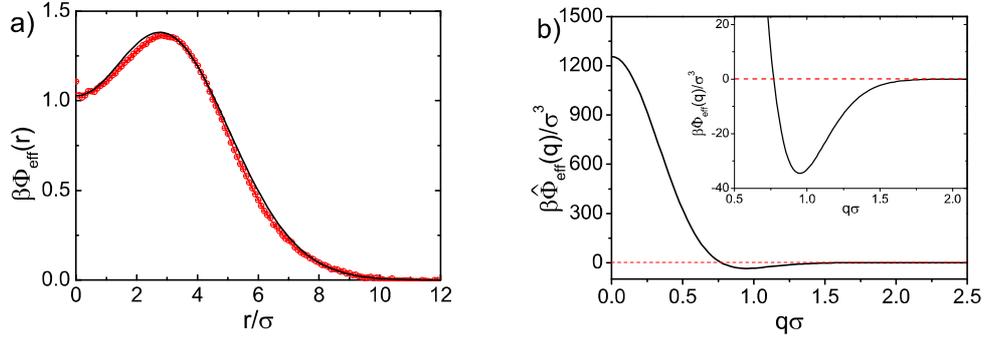

Figure 1: a) The effective pair-potential $\Phi_{\text{eff}}(r)$ for amphiphilic dendrimers as obtained from MC simulations (red circles) and a fit to the difference of two Gaussian potentials (black line); and b) its Fourier transform $\hat{\Phi}_{\text{eff}}(q)$, with the inset showing a zoom of the region around its most negative value, $-|\hat{\Phi}_{\text{eff}}(q_*)|$.

Table 2: Parameters for the four systems considered in this study, with $\rho\sigma^3$ being the dendrimer number density and L the size of the simulation box.

| system | $\rho\sigma^3$ | $L/\sigma$ |
|---|---|---|
| A | 0.010 | 29.24 |
| B | 0.015 | 25.54 |
| C | 0.020 | 23.20 |
| D | 0.025 | 21.54 |

# 3  Cluster assignment algorithm

Although the term clustering has been mentioned several times above, it is not immediately evident how to define a cluster in a proper way and under which condition particles are considered to belong to the same cluster. In the present case, the problem is particularly delicate: first, as the density is being increased, dendrimers approach each other and assemble into groups, as, for instance, can be found in the process of nucleation; in addition, dendrimers can even penetrate each other to an extent where the centers of mass of two overlapping particles lie on top of each other. For this reason, a criterion is required that allows to sort the dendrimers into clusters.

To this end, a minimum distance $d_{min}$ is introduced, which acts as a standard length to decide whether two dendrimers are considered to be part of the same cluster or not. In cluster forming solids, e.g. the multiple occupancy crystals formed by GEM-4 particles,[5] one usually finds the individual particles to gather within roughly spherically shaped groups. Still, even for this case the identification to which cluster a particular particle belongs is not trivial.[32]

Extensive visual inspection of configurations obtained from our simulations confirm that many clusters do seem to have this shape. However, similar to what is observed in GEM-4 fluids, also aspherical aggregates are observed. In addition, more extended clusters are found that are formed by two or more groups of particles that are connected by intermediate, bridge-like structures of dendrimers. Such bridges can be formed in the process of exchanging particles between neighboring clusters, as well as from the natural fluctuations and rearrangements of dendrimers within clusters. For this reason, a simple algorithm based only on the mutual distance between dendrimers is not sufficient.

We have modified the usual type of cluster algorithm based on the relative particle distances in order to separate at least the most simple types of extended clusters in their constituting components. Such aggregates contain bridges formed by single, almost linear, chains of dendrimers that connect the different sub-clusters. This separation can be achieved by filtering out isolated dendrimers or dendrimers with a low connectivity in the following fashion:

1. For each dendrimer, the list of $N_{NN}$ nearest neighbors is determined, for which the centers of mass lie within the distance $d_{min}$ of the center of mass of the dendrimer that is considered.

2. If for a dendrimer $N_{NN} = 0$ or if $N_{NN} = 1$ and the only neighbor is found at a distance larger than $0.5 \times d_{min}$, the dendrimer is counted as a single dendrimer cluster.

3. If $N_{NN} = 2$ and the angle $\alpha$ between the vectors to both neighbors satisfies $\alpha > \frac{\pi}{2}$ and both are are found at a distance larger than $0.5 \times d_{min}$, the dendrimer is also counted as a single dendrimer cluster. These are dendrimers that are located within an almost linear chain of dendrimers between different clusters.

4. Since some dendrimers are counted as individual clusters, the number of neighbors of the remaining dendrimers might have changed. Therefore, step 2 and 3 need to be iterated until no more single-particle clusters are found.

5. The remaining list of dendrimers is sorted into clusters according to the neighboring particles.

For all results presented here, $d_{min} = 3\sigma$ has been used, a value which is somewhat smaller than the zero-density radius of gyration of the dendrimers,[1] and also smaller than the typical distance $4\sigma$ that corresponds to the first minimum in the radial distribution function of the centers of mass of the dendrimers (see Figure 6). This ensures that dendrimers that are sorted into the same cluster are close in terms of the size of an individual dendrimer as well as with respect to the length scale of their global structure.

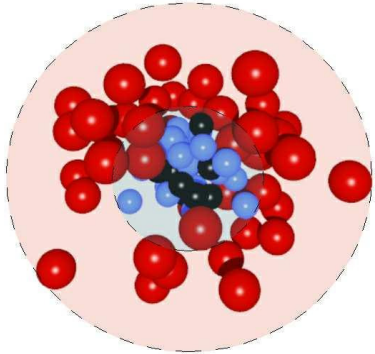

Figure 2: A typical cluster of dendrimers, having an occupancy number $N_{occ} = 6$ as found in simulations of system C. Core monomers are black (central monomers) and blue (first generation), respectively, shell monomers are red. For clarity, bonds are not shown.

A typical example of a cluster of dendrimers with an occupancy number $N_{occ} = 6$ and taken from system C is shown in Figure 2. The core particles are found to be closely grouped together in the central region, whereas the shell monomers are more loosely distributed outside. It should be noted that the simple cluster algorithm described above can easily deal with this type of roughly spherical clusters, but that it is not capable of resolving the problem of some of the more complex aggregates. For instance, strongly elongated structures or dumbbell-shaped aggregates will be still classified as single clusters. The same is true when two spherical clusters are connected by a more complex type of bridge, e.g. a double chain of dendrimers.

## 4 Spontaneously assembled clusters

To give an impression of the four systems investigated in this study, a snapshot of equilibrated configurations for each of the densities is shown in Figure 3. Whereas in the lowest density, system A, no clear structure can be observed, the higher density systems indicate an increasingly evident presence of clusters. Since in all cases the simulations were started from independent, randomly distributed dendrimers at the given density, the formation of clusters occurs spontaneously. This confirms that the prediction of this behavior by theoretical works and simulations of coarse-grained, effective particles is correct.[9]

The cluster-participating dendrimers in these systems are not arrested within their clusters; to the contrary, they are mobile. Within the equilibrated simulation runs, dendrimers travel over distances of several box-lengths and clusters are continuously formed, merged and broken up into smaller ones. The latter processes can easily be verified by measuring the density of clusters, i.e. the amount of clusters existing at any given moment divided by the volume of the simulation box, as obtained by using the cluster algorithm outlined above. Figure 4 shows the average density of clusters, using only the number of clusters consisting of two or more dendrimers, as a function of the overall dendrimer density for equilibrated systems. Also the spread in this number is indicated, which was obtained by using the minimum and maximum number of clusters at every given density within each of those simulation runs to determine the density of clusters. For low densities, the system is fluid-like and has a low cluster density, since only few clusters are detected which are basically due to local fluctuations and which lack a strong cohesion. On increasing the dendrimer density, the cluster density grows and appears to reach a maximum at about $\rho\sigma^3 = 0.015$, after which it starts to decrease again. The latter decrease is a consequence of the fixed number of total dendrimers and their tendency to form bigger clusters at higher densities. In addition, it should be kept into consideration that the characterization of an aggregate as cluster does depend on the specifics of the cluster-assignment algorithm.

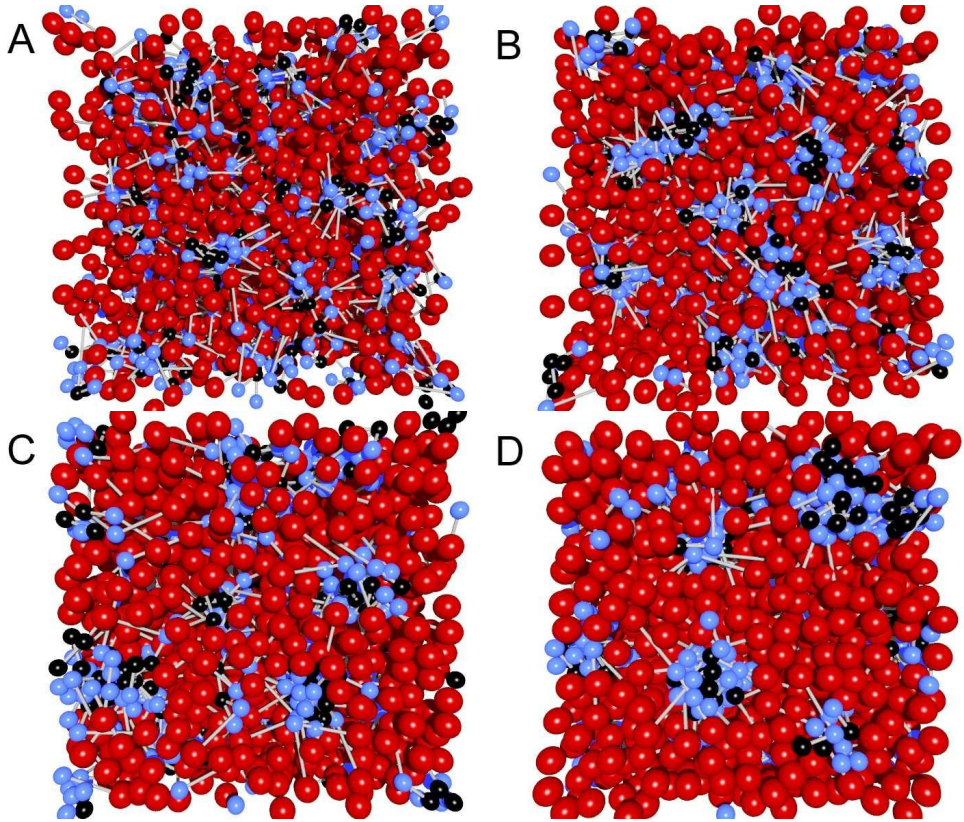

Figure 3: Snapshots of systems A-D. Colors as in Figure 2 and bonds are indicated in gray.

To analyze the process of dendrimers hopping between clusters in more detail, the cluster size distributions have been calculated. The results for systems A to D are shown in Figure 5, where the normalized probability size-distributions $P(N_{occ})$ are shown as function of the cluster size $N_{occ}$. For the low density $\rho\sigma^3 = 0.01$ of system A, only a few, relatively small, clusters are found. At the higher density $\rho\sigma^3 = 0.015$, system B, one can already observe that an increasing number of clusters is formed of up to 12 dendrimers. Apart from the single particle clusters, the most frequently found clusters are still small with an $N_{occ}$ in the range of 3 to 5.

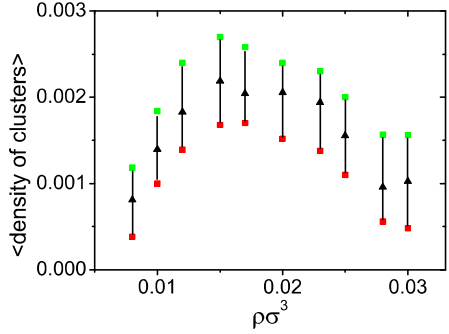

Figure 4: The average density of clusters as function of the dendrimer density (triangles). The minimum and maximum value of the cluster density within each simulation run are also indicated. Single particles clusters were not considered for the determination of the cluster density.

On increasing the density to $\rho\sigma^3 = 0.02$, system C, the number of larger clusters increases and their preferred size shifts to the range 7–10. At the same time, the number of smaller clusters with 2–6 dendrimers diminishes. For the highest density considered here, this trend progresses. The most frequently found clusters now contain about 10 dendrimers. The probability to find clusters of twice that size is almost as high, and even clusters three times as large are found. This considerable tail in the distribution of much larger cluster sizes, of which the onset could already be seen in system C, is caused by extended clusters, i.e., two or three clusters of average size, but connected by a string of dendrimers that cannot be resolved by the simple algorithm outlined above. In addition, the system contains only 250 dendrimers and is therefore on the level of clusters rather small, hence finite size effects might play a role.

It is clear that a coarse graining of the system by approximating the dendrimers by spheres interacting according to the zero-density limit effective interaction cannot be perfect. By using the pair-interaction only, we neglect all many-body effects, and hence some deviations from monomer resolved simulations should be expected. Such corrections are necessary even at the lowest density considered in system A, $\rho\sigma^3 = 0.01$, since it is already quite higher than the overlap concentration of $\rho_* \cong 3/(4\pi R_g^3) \cong 0.006$.

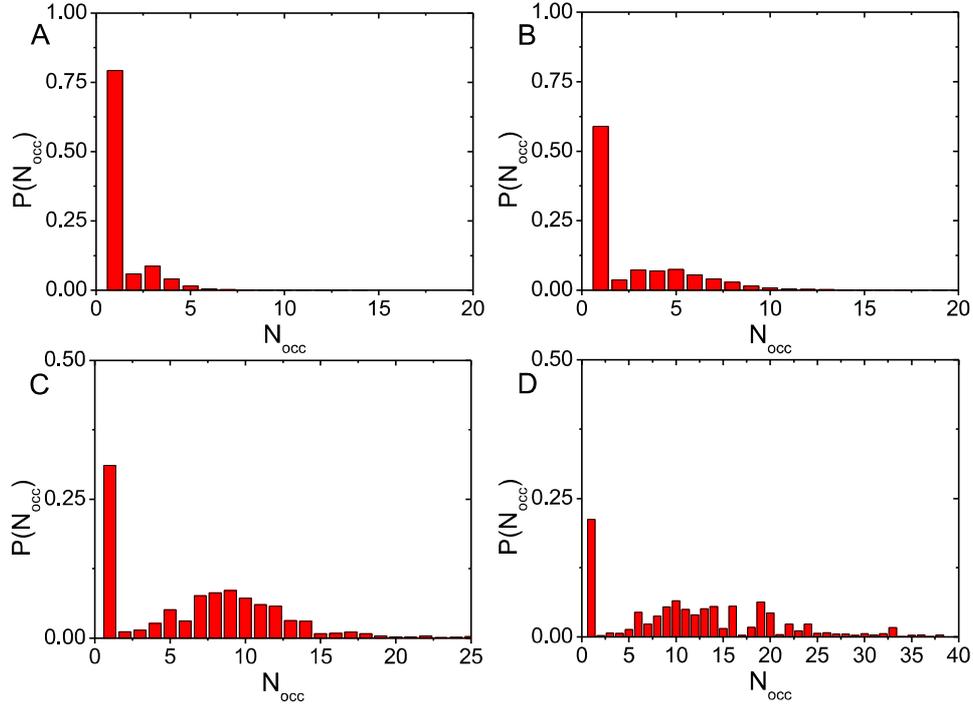

Figure 5: The normalized probability distribution of cluster sizes, $P(N_{occ})$, for the systems A-D.

To this end, we discuss below the comparison between key results obtained by two different approaches: one is the monomer-resolved simulation and the other a simulation of point particles interacting by means of the infinite-dilution effective pair potential $\Phi(r)$ only. In particular, we consider the radial distribution function of the centers of mass of the dendrimers, $g(r)$, as is illustrated in Figure 6. In either approach, the local maxima found at $g(0)$ indicate that dendrimers can get arbitrarily close and have the propensity to cluster. It is in this limit of close approach between dendrimers that the coarse grained results deviate somewhat from the monomer resolved results and underestimate the strength of clustering. This is hardly surprising since the increasing density will affect the effective pair-interaction especially in this range. The first correction term, i.e., the three-body interaction, is evidently attractive, since the monomer-based approach yields a higher value of $g(0)$ than the effective one does. However, upon increasing the concentration further, the trends reverts itself. At the highest density considered, there is a significant discrepancy between the two approaches. Although cluster formation survives in the monomer-resolved approach, it is much less pronounced than in the effective one, based on pair potentials. For this density, which is already higher than the one predicted for the instability of the fluid and the freezing into cluster solids,[4] the monomer-resolved simulations hint towards a zone around a given cluster, which is depleted from other dendrimers. This fact, in itself, is at least an indication that the system attempts a structural relaxation and possibly tends to form an ordered structure on the level of clusters. It should be noted, however, that the system of 250 dendrimers with a mean cluster-size $N_{occ} \approx 10$ is relatively small, and that a full structural relaxation will hardly be possible due to the periodic boundary conditions.

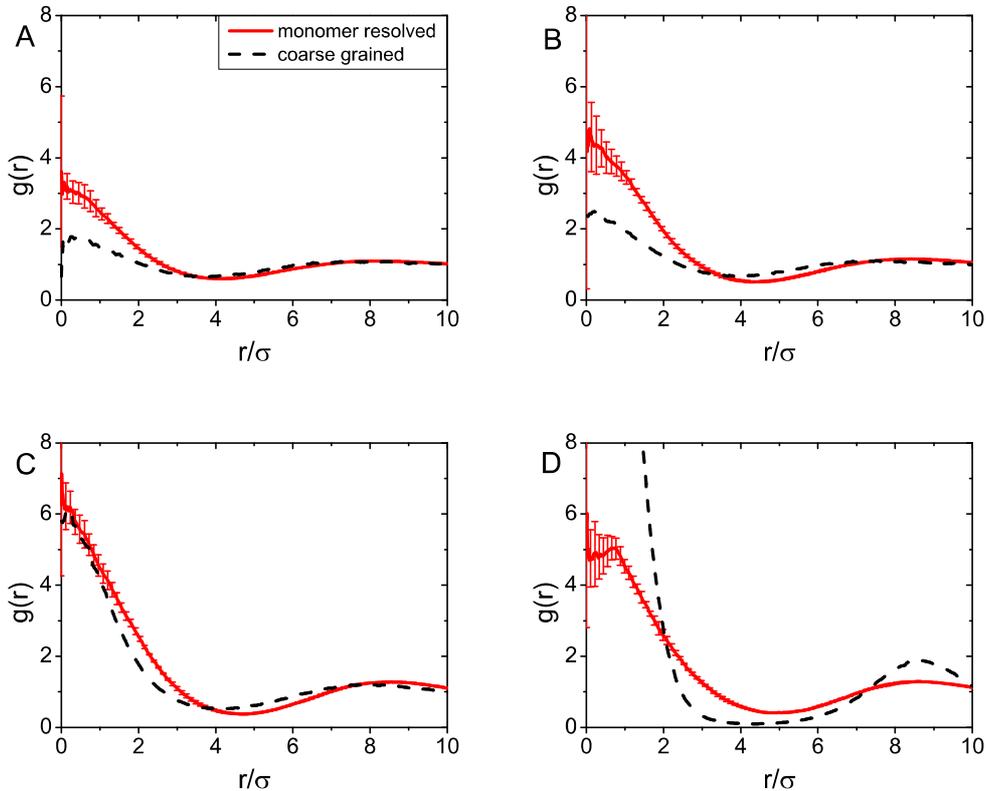

Figure 6: Radial distribution function of the centers of mass of the dendrimers for systems A-D. Red line: monomer resolved, black dashes: coarse grained.

The radial distribution function of the centers of mass of the clusters, $g_{cl}(r)$, is shown for system

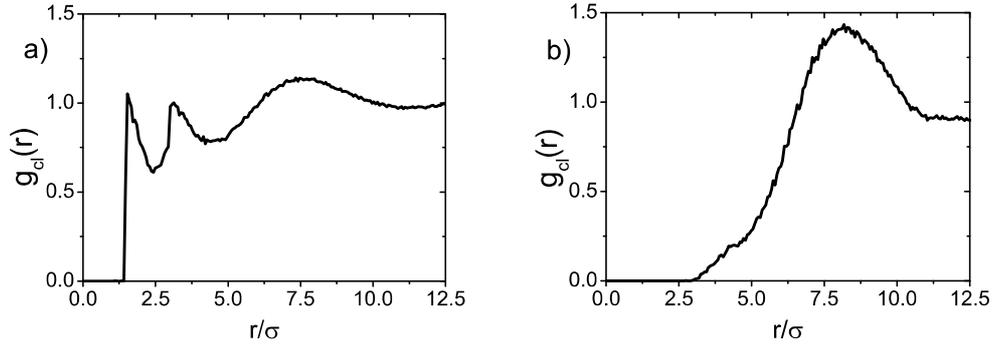

Figure 7: Radial distribution function for the centers of mass of clusters for system B. a) Single particle clusters are included, and b) excluded.

B in Figure 7. The two peaks at $r \approx 1.5\sigma$ and $r \approx 3\sigma$ are due to single dendrimers that approach either each other or a larger cluster, and which are assigned by the algorithm to different clusters. This can be seen from the right panel in the same figure, where we show the radial distribution function as obtained when those single particle clusters are excluded from the analysis. Now, the first two peaks which are a signature of the processes described above, disappear. The remaining cluster correlation turns out to be more or less independent of the choice of $d_{min}$.

## 5 Internal cluster structure

Measurements of the internal dendrimer structure by means of the density profiles of core and shell monomers with respect to the center of mass of the dendrimer reveal that these functions are hardly affected by the increased overall density of the system. They remain very similar to those of the infinite dilution regime by Mladek et al.[9] and are therefore not presented here. Also the mean radius of gyration in the observed systems remains close to the value $R_g \approx 3.36\sigma$ of an isolated dendrimer. This does, however, not necessarily exclude the possibility of other deformations, such as the stretching of internal bonds due to crowding of the core monomers, or shrinking of dendrimers caused by the external pressure exerted by other clusters.

In order to probe the internal structure of the clusters, density profiles of both types of monomers

(core and shell) and of the centers of mass of the dendrimers are measured with respect to the center of the cluster under study. Hereby the contributions are separated in those monomers and dendrimers that belong to the cluster ("member"), as well as those stemming from neighboring clusters ("non-member"). Figure 8 shows the results for systems A-D for the most frequent cluster sizes.

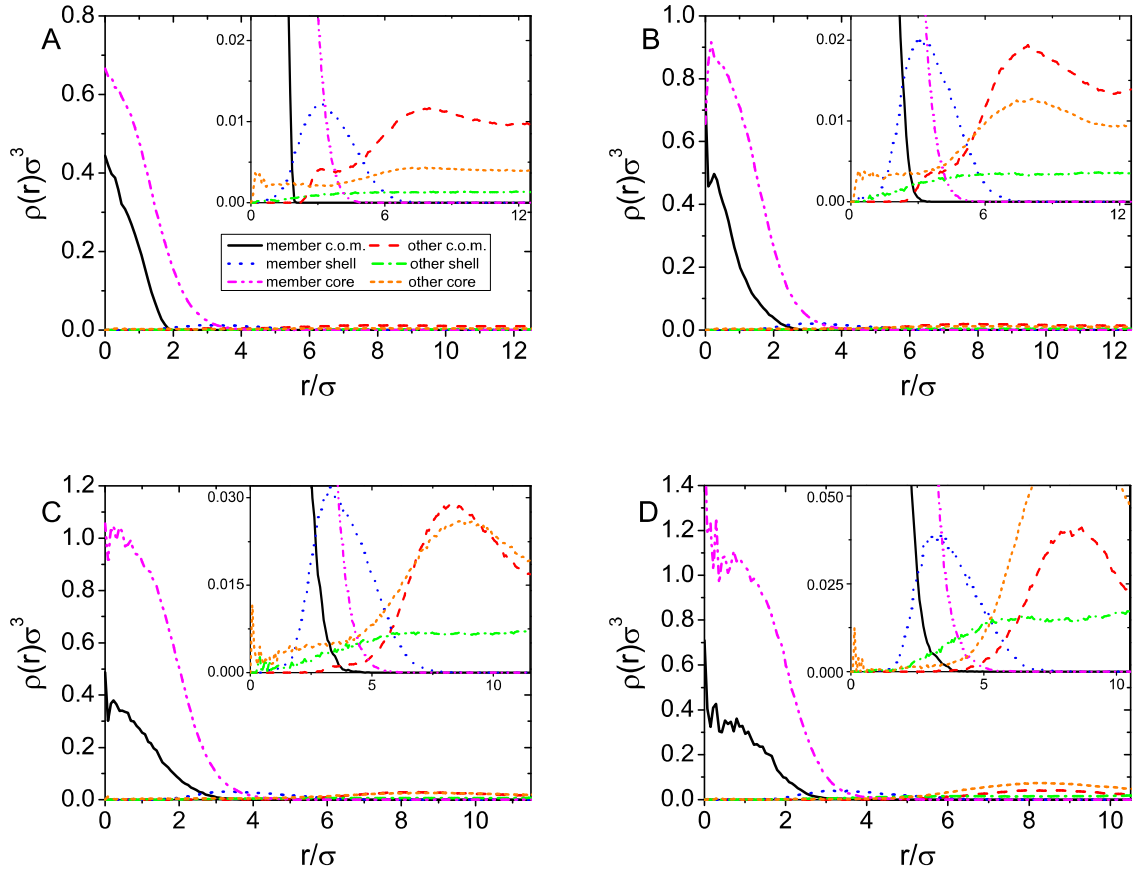

Figure 8: Internal radial density profiles of monomers and center of mass (c.o.m) of dendrimers for both cluster members and non-members ("other"). The number of dendrimers per cluster are as follows: system A) $N_{occ} = 3$, B) $N_{occ} = 5$, C) $N_{occ} = 9$, and D) $N_{occ} = 10$. The insets show zooms within regions of the density profiles, in which the latter feature considerable substructure.

The profiles of the monomers belonging to the cluster, i.e. the member ones, are qualitatively very similar to those of the monomers inside a dendrimer. This can be understood from the fact that the amphiphilicity causes a spatial segregation between core and shell monomers. Hence, the core monomers are essentially building the core of the clusters, whereas the shell monomers form a surrounding cloud, as is also illustrated by the snapshots in Figures 2 and 3. In addition, the shape of these profiles depends only weakly on the number of dendrimers within the cluster.

Surprisingly, also numerous isolated monomers of dendrimers that do not belong to the cluster under study can be found in the central region of this cluster. To understand this behavior, we additionally plot the density profile of the centers of mass of the non-member dendrimers in Figure 8 and consider the first peak in this distribution. This peak mostly stems from dendrimers that are still close to the cluster under study, but have been sorted out by the cluster algorithm and are therefore either single dendrimer clusters, or dendrimers that are part of a simple chain-like bridge. Their core monomers, which on average are evenly distributed around the center of mass of the dendrimer to which they belong, extend into the core of the cluster under study, which acts as an attractive basin due to the amphiphilic interactions. For the same reason, the shell monomers of the non-member dendrimers are expelled from that region, even though this area in principle lies within their reach. The second, much larger peak in the density distribution of the centers of mass of non-member dendrimers corresponds to the average distance between neighboring clusters. The corresponding core monomers remain in general outside the cluster under study. By contrast, the clouds of the shell monomers of neighboring clusters have a substantial overlap. Increasing the density as we go from system A to D enhances the cohesion in the clusters and results in less and less isolated dendrimers. Consequently, the non-member monomers get more and more expelled from the central cluster in the systems B and C. For the highest density of system D, the compactness of the cluster core and its surrounding cloud prevent almost all non-member core monomers to penetrate the central region.

The profiles of the internal structure of the clusters just shown in Figure 8 correspond to the most frequent cluster-size for each of the system densities. Consequently, the occupation numbers $N_{occ}$ are different, which makes it difficult to compare the functions. To this end, Figure 9 shows the same profiles but now normalized by the occupation number $N_{occ}$. This reveals that the central peak in the core distributions decreases with increasing density and at the same time becomes somewhat wider. These trends can easily be explained by the crowding of the core monomers taking place in the center of the cluster by the increased number of dendrimers. By contrast, the shell-monomers, which have a much weaker mutual interaction, are hardly affected by the increase in density and are more evenly distributed in space. The fact that not only the height of the peak in the shell monomer distributions, but also its distance from the center of the cluster remains unchanged can be understood from the fact that the radius of gyration of the dendrimers is almost constant in the density range considered here. The core monomers of the non-member dendrimers are also found at nearly the same distance for all densities considered, which indicates a preferential distance between clusters that is almost independent of the density. A closer examination of the monomer density profiles for clusters consisting of the same number of dendrimers $N_{occ} = 5$ as shown in Figure 10 reveals that the increasing density of the system compactifies the cluster. Both the core and monomer profiles are somewhat more concentrated. It should be realized, however, that for both systems A and D, this occupation number has a rather low probability.

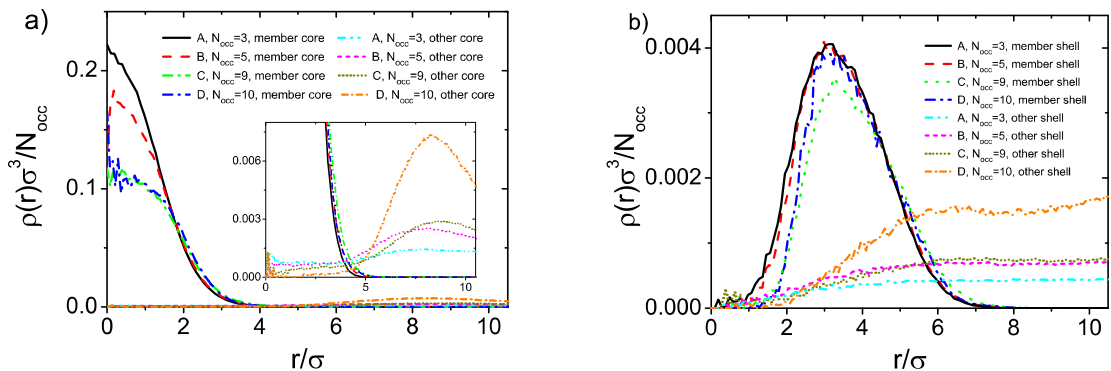

Figure 9: Cluster internal density profile normalized by the number of dendrimers per cluster for a) core monomers and b) shell monomers.

By contrast, if one examines the profiles at the same density for different cluster sizes as can be seen in Figure 11, an increased occupation number $N_{occ}$ results in a spatially more extended cluster. This is not unexpected, since the increase in the number of dendrimers sitting on top of each other in a cluster results in an increase of the internal repulsive forces of the cluster. This effect is probably enhanced even more by the fact that in the case of the largest occupation numbers, the cluster size already exceeds the most frequent size, and hence the spherical shape of the cluster becomes distorted as dendrimers try to diffuse to other, less occupied clusters.

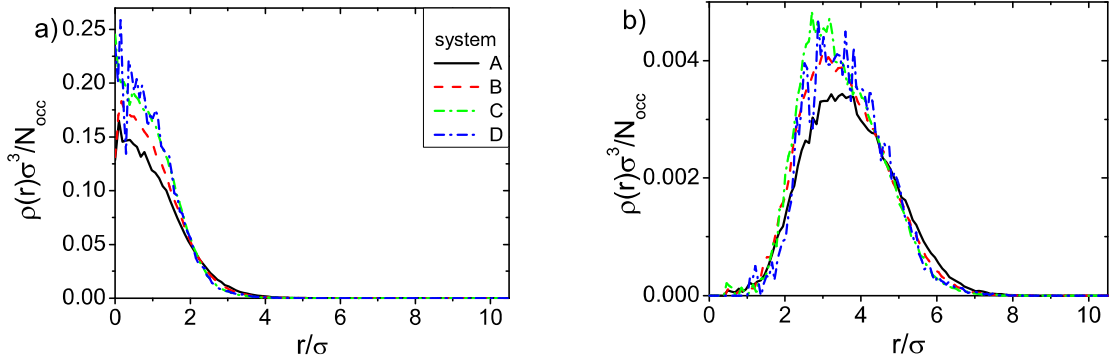

Figure 10: Cluster internal density profile of a) member core monomers and b) member shell monomers for fixed occupation number $N_{occ} = 5$ at systems A-D. The curves are normalized by occupation number.

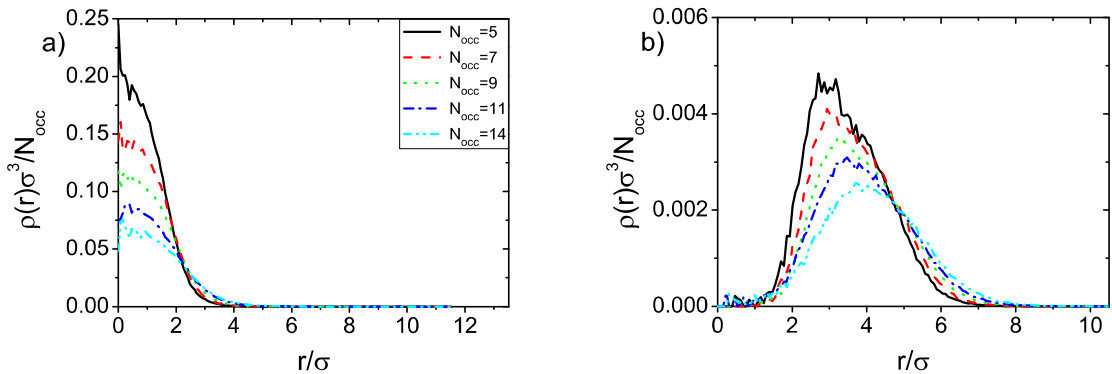

Figure 11: Cluster internal density profile for a) member cores and b) member shells, respectively, for different cluster occupation numbers in System C. Data normalized by respective occupation number.

# 6  Conclusions

With the aid of Monte Carlo simulations we have investigated the behavior of complex liquids of amphiphilic dendrimers, in particular their potential to cluster-formation as was predicted theoretically. This prediction was made on a coarse-grained level, for which the dendrimers are approximated by soft spheres that interact with the effective pair-interaction of two isolated dendrimers. Here, we have shown that this behavior to form clusters is confirmed by a description on the monomer level. To this end, random configurations of dendrimers at various densities have been generated and allowed to equilibrate. During this process, the systems spontaneously creates clusters which can already be visually observed from snapshots, and which can more rigorously be identified by a relatively simple cluster algorithm. While these clusters are found to be roughly spherical for low densities, it should be noted that at higher densities they can be connected by bridges of intermediate dendrimers. These bridges are a signature of an active exchange of dendrimers between different clusters, which can also be seen from the fact that individual dendrimers travel over multiple box-lengths within the duration of the simulation.

The radial distribution functions of the centers of mass of the dendrimers corroborate that the coarse-grained level description is qualitatively correct and confirms that these systems form a cluster fluid. However, for densities around the one for which the freezing into cluster crystals has been predicted, a significant quantitative discrepancy in the radial distribution functions is observed. Consequently, monomer resolved simulations are probably required for a quantitative analysis of dendrimer systems at such high densities. Furthermore, we have presented a detailed analysis of the monomer as well as dendrimer distributions within spontaneously formed cluster structures, which gives interesting new insights into the complex behavior of thus assembled macromolecules and might pave the way for an advanced analysis of such kind of complex fluids up to freezing. With this, the presented work delivers a confirmation that amphiphilic dendrimers are eligible molecules for further theoretical and computational studies as well as a possible direct experimental realization of the fascinating novel phenomenon of macromolecular clustering.

# 7  Acknowledgments

B.M.M. acknowledges funding from the EU via FP7-PEOPLE-IEF-2008 No. 236663. G.K. acknowledges financial support by the Austrian Science Foundation (FWF) under Project No. P19890-N16. This work has received support by the Marie Curie Training Network ITN-COMPLOIDS, FP7-PEOPLE-ITN-2008 No. 234810.